# In-Plane and Out-of-Plane Excitonic Coupling in 2D Molecular Crystals

Dogyeong Kim[1], Sol Lee[2,3], Jiwon Park[1], Jinho Lee[1], Hee Cheul Choi[1], Kwanpyo Kim[2,3] and Sunmin Ryu[1]*

[1]Department of Chemistry, Pohang University of Science and Technology (POSTECH), Pohang, Gyeongbuk 37673, Korea

[2]Department of Physics, Yonsei University, Seoul, 03722, Korea

[3]Center for Nanomedicine, Institute for Basic Science (IBS), Seoul, 03722, Korea

*E-mail: sunryu@postech.ac.kr

**ABSTRACT**

Understanding the nature of molecular excitons in low-dimensional molecular solids is of paramount importance in fundamental photophysics and various applications such as energy harvesting, switching electronics and display devices. Despite this, the spatial evolution of molecular excitons and their transition dipoles have not been captured in the precision of molecular length scales. Here we show in-plane and out-of-plane excitonic evolution in quasilayered two-dimensional (2D) perylene-3, 4, 9, 10-tetracarboxylic dianhydride (PTCDA) crystals assembly-grown on hexagonal boron nitride (BN) crystals. Complete lattice constants with orientations of two herringbone-configured basis molecules are determined with polarization-resolved spectroscopy and electron diffraction methods. In the truly 2D limit of single layers, two Frenkel emissions Davydov-split by Kasha-type intralayer coupling exhibit energy inversion with decreasing temperature, which enhances excitonic coherence. As the thickness increases, the transition dipole moments of newly emerging charge transfer excitons are reoriented because of mixing with the Frenkel states. The current spatial anatomy of 2D molecular excitons will inspire a deeper understanding and groundbreaking applications of low-dimensional molecular systems.

hydride (PTCDA)

**Introduction**

Excitons in molecular solids mediate efficient light harvesting in photosynthesis complexes[1] and important organo-electronic applications such as photovoltaics[2] and light-emitting diodes.[3] They also present a myriad of intriguing photophysical phenomena such as Davydov splitting (DS)[4, 5] associated with J/H aggregates,[6] coherence-induced superradiance,[7] mixing of Frenkel and charge transfer [8] excitons,[9] and singlet fissions.[10] Despite their small footprints spanning a few unit cells, molecular excitons are greatly affected by intermolecular interactions through Kasha-type Coulombic[11] and intermolecular CT[12] couplings. Geometric arrangements are crucial in both interactions in that the former is governed by the long-range coupling among transition dipoles,[13] whereas the latter is by the short-range orbital overlap.[14] Notably, the CT coupling is ubiquitous in π-stacked systems, differentiates the J/H-aggregate behaviors,[15] limits the coherence of Frenkel excitons (FE)[16, 17] and facilitates the singlet fission.[18] Despite many experimental breakthroughs and progress,[19] however, it remains unclear how the evolution of Frenkel-CT mixing can be revealed by other than differing energetics.[20-22] By exploiting the capability of polarized spectroscopy that directly maps the transition dipoles, more direct proof for the mixing and its spatial progression should be within experimental reach.

For such an exploration, two-dimensional molecular crystals (2DMCs) can be an ideal system.[23, 24] In principle, the above interactions can be tuned by varying intermolecular arrangements, dielectric environments and the extent of coupled units. In bulk crystals, however, the two former parameters are thermodynamically fixed and can hardly be modified except for polymorphic transitions. Whereas the number of molecular units under excitonic coupling can be purposefully varied by forming oligomers with chemical bonds,[25] it is not applicable to the molecular crystals formed via van der Waals (vdW) interactions, either. In contrast, variable-thickness 2DMCs epitaxially grown on crystalline templates can provide crucial controls over the geometric factors. Most of all, the crystalline structure of 2DMCs is affected by the templates, which may enable customizing intermolecular geometry.[24, 26] Monolayer-resolved thickness control will allow fine-tuning of the spatial extent of excitonic coupling[27] and the degree of dielectric screening.[28] The distinction between in-plane and out-of-plane couplings may also be made by varying thickness approaching the truly 2D limit.

In this work, we directly reveal the in-plane and out-of-plane excitonic evolutions of 2D PTCDA (perylene-3,4,9,10-tetracarboxylic dianhydride) crystals grown on hexagonal BN (hBN) substrates in an unprecedented manner using polarized photoluminescence (PL) and absorption spectroscopies. The crystalline structure of PTCDA and its epitaxial registry with hBN were revealed by electron diffractions. The excitons confined within monolayer PTCDA exhibited enhanced in-plane coherence and superradiance at cryogenic temperatures, which was supported by the inversion in DS. We also show that PL from CT excitons is dominant in multilayer PTCDA and their transition dipoles are drastically reoriented with increasing thickness as a result of varying CT contribution in Frenkel-CT mixing. The anatomy of the excitonic couplings in 2D PTCDA shown in this work can be applied to other 2DMCs and serve as a touchstone for the theoretical modeling of Frenkel-CT interactions.

**Results and discussion**

**Structural and optical characterization of 2D PTCDA crystals.** Single and few-layer 2D

PTCDA crystals ($nL_P$) shown in Fig. 1a were grown on few-layer hBN ($nL_{BN}$) by physical vapor deposition (PVD) at elevated temperatures. The AFM height image (Fig. 1a) and profile (Fig. 1b) showed a flat layered structure with an interlayer spacing of 0.31 ± 0.08 nm (more examples with optical micrographs are given in Supplementary Figure 1). As described in Methods, crystalline substrates for growth and excess thermal energy[23, 24] were crucial in obtaining ordered structures. The crystalline nature of the samples was revealed by electron diffraction, as shown in the selected area electron diffraction (SAED) image of $2L_G/1L_P/nL_{BN}$ (Fig. 1c), where the encapsulation with bilayer graphene ($2L_G$) led to greatly enhanced stability against electron-induced damages (see Methods). Analysis of multiple samples verified that the rectangular unit cell determined by the diffraction method corresponds to (102)-plane unit cell of bulk crystals.[29] Then, *m* and *n* spanning the 2D unit cell (Fig. 1d) are equivalent respectively to *b* and $[\bar{2}01]$ of bulk: *m* = 1.256 ± 0.018 (1.220 ± 0.030) nm, *n* = 1.971 ± 0.017 (1.972 ± 0.008) nm for $1L_P$ ($3L_P$). As summarized in Supplementary Table 1, *m* of $1L_P$ was ~6% longer than that of bulk,[30] but the deviation was reduced for $3L_P$, which suggests convergence toward the bulk value with increasing thickness. In addition, the fact that interlayer spacing in 2D PTCDA is very close to that of bulk[30] corroborates that $1L_P$ corresponds to a single (102) plane.[29] As shown below, DS in PL spectra indicated that two basis molecules are arranged in a herringbone configuration as in bulk crystals.[30] The orientation of each basis molecule within the unit cell could be determined by polarization-resolved absorption spectra as described in Supplementary Figure 2. The simulated diffraction pattern in Supplementary Figure 3 also showed that 2D PTCDA is closer to the α polymorph, preferred for higher growth temperature, rather than β[31] (see Supplementary Figure 4 for their structural difference). Note that the unit cells in two neighboring layers are displaced against each other along *n* as depicted in Fig. 1d. The formation of the 2D crystalline layered structure is assisted by the interplanar π-π interaction[32, 33] and in-plane hydrogen bonding.[34] We also note there is a specific orientational registry between nL$_P$ and underlying hBN: the ***n*** axis of 1L$_P$ and 3L$_P$ formed 10.6 ± 0.1° with the armchair direction of hBN (Supplementary Figure 3h). This indicates that the crystalline nature of 2D PTCDA is supported by substrate-mediated epitaxial growth. In contrast, graphene allows two different rotational polytypes (Supplementary Figures 3b and 3c). More restrictive binding on hBN than graphene was also reported for tetracene molecules.[24]

In addition to the topographic measurements, the average thickness of PTCDA samples could be determined by optical contrast (Methods), which is proportional to the degree of absorption for thin samples supported on transparent substrates.[35] Figure 1e showed a good resolution for thickness and linearity up to 6 layers. The spectral changes in the room-temperature PL spectra (Fig. 1f) were largely consistent with previous studies on bulk crystals and thin films:[32, 33] whereas monolayer samples present localized FEs with a substantial vibronic progression, thicker samples are dominated by CT excitons denoted as CT$_{high}$ and CT$_{low}$, which are coupled with lattice phonons to varying degrees.[32, 33] CT excitons have been modeled to reside in 1D molecular stacks under the assumption that the in-plane excitonic coupling is negligible compared to the interplanar π-π interactions.[9, 20, 36] Although the low-energy PL peaks are mostly of the CT-exciton character,[9, 32] the exact nature and assignment of each peak is far from reaching a consensus.[32, 33] We note that a systematic study is required to relate the current CT$_{high}$ and CT$_{low}$ of few-layer PTCDA with those found for bulk samples.

**Davydov splitting and in-plane excitonic coupling of 1L PTCDA crystals.** The bulk-like rectangular unit cell determined by the diffraction measurements suggests that each unit cell contains two basis molecules hydrogen-bonded in a herringbone configuration of α polymorph[30] (Supplementary Figure 3). According to Kasha's dimer model,[11] the FE state of a two-basis system undergoes DS and splits into two states in general. As depicted in Fig. 2a, the Coulomb interaction ($J_{Coul}$) converts the degenerate excited states of a non-interacting dimer (denoted as $\varphi_I^*$ and $\varphi_{II}^*$) into the lower and upper DS states, denoted as LDS and UDS, respectively. The two DS states can be described as the symmetric and antisymmetric superposition of $\varphi_I^*$ and $\varphi_{II}^*$. Under a point-dipole approximation, the split energy ($E_{DS}$) corresponding to $2J_{Coul}$ will be governed by mutual orientations and displacement between the two transition dipole moments that are represented by the red (blue) arrow pairs in Fig. 2a. Their transition dipole moments ($\mu_{LDS}$ and $\mu_{UDS}$) are expected to be orthogonal to each other, no matter what angle the transition dipole moments of the two bases ($\mu_{\varphi I}$ and $\mu_{\varphi II}$) form.[5] To determine the orientations of the DS dipoles, we obtained polarized PL spectra of $1L_P$ in a parallel configuration by rotating excitation polarization (see Supplementary Figure 5 for a complete set of spectra). In Fig. 2b presenting LDS and UDS-polarized spectra, the 0-0 transition led to two prominent DS peaks with a split energy ($E_{DS}$) of 36 meV. Their polar intensity graph in Fig. 2c verified their orthogonality (90.2 ± 1.3 deg for multiple samples). UDS still showed significant signals at the intensity minimum unlike LDS, the origin of which is not clear. The DS of 0-1 and 0-2 peaks could not be discerned because of their large spectral widths and overlap. Notably, the DS has not yet been observed in any form of PTCDA solids directly by emission or absorption spectroscopy. In addition to the lacking availability of single crystals of sufficient quality and size, dominant CT emission[37] and extremely high absorption[38] impede the conventional polarized measurements, respectively. The only experimental $E_{DS}$, 37 meV for α-PTCDA, was determined by ellipsometry.[38]

Correlated measurements of electron diffraction and polarized PL (Supplementary Figure 2) allowed us to determine the orientation of $\mu_{LDS}$ and $\mu_{UDS}$ with respect to the crystallographic coordinates of 2D PTCDA. As shown in Fig. 2e for $1L_P$, $\mu_{LDS}$ was found to make an angle of 21.5° with the *n* axis. We note that this information can be used in determining the molecular orientation in 2D PTCDA because molecular transition dipoles are aligned along the long molecular axis.[17] Assuming that $\mu_{\varphi I}$ and $\mu_{\varphi II}$ are identical in their magnitudes, $\mu_{LDS}$ ($\mu_{UDS}$) is parallel (perpendicular) to the bisector of the angle ($\theta_{I/II}$) formed by $\mu_{\varphi I}$ and $\mu_{\varphi II}$ (Fig. 2e). As $\theta_{I/II}$ governs the relative magnitude of the DS dipoles, it could be determined by comparing the absorption by LDS and UDS (see Supplementary Figure 2 for details). The scheme in Fig. 2e depicts the molecular arrangement of $1L_P$ determined by the diffraction, polarized PL and absorption measurements. The fact that $\theta_{I/II}$ = 89.2° indicates that the two bases are aligned almost perpendicularly to each other unlike bulk crystals with $\theta_{I/II}$ = 96°.[29]

In Fig. 3, we show a distinctive temperature dependence of the DS peaks of $1L_P$. Whereas the emission from UDS (LDS) of J (H) dimers is dipole-forbidden,[11] both DS transitions of oblique dimers found in 2D PTCDA and tetracene[24] are allowed. Because of rapid internal conversion,[39] the emission from UDS is insignificant unless $E_{DS}$ is close to or smaller than the thermal energy like PTCDA (Fig. 2a). For example, 2D tetracene showed negligible UDS intensity because of its large $E_{DS}$ (~95 meV) which hinders the thermal population of UDS.[24] The two sets of polarized PL spectra of $1L_P$ (Fig. 3a and 3b) clearly show that LDS and UDS redshift with decreasing temperature. Remarkably, the rate of redshift is 2.7 times larger for UDS than LDS as shown in Fig. 3c (see Supplementary Figure 6 for more data and Supplementary Figure 7 for complete spectra), which leads to the inversions in the peak energies (Supplementary Figure 7) and intensity ratio (Fig. 3c and 3d). The spectral narrowing (Fig. 3e) accompanied by the increase of their intensities suggests that their excitonic decay is mainly governed by superradiance,[40] an accelerated spontaneous emission from coherently coupled[41] or delocalized[42] excitons. The degree of delocalization will be discussed later using the molecular coherence number, $N_c$ (Fig. 3f), which was extracted from the linewidth (σ).[43]

The temperature dependence of both DS peaks in Fig. 3 provides a crucial insight into the nature of excitonic coupling in $1L_P$. According to Kasha's exciton model,[11] the excited state of a herringbone-configured PTCDA dimer splits into LDS and UDS (Fig. 4a), and their average energy is slightly lower than that of the excited monomer. The former is induced by the Coulombic interaction between two transition dipoles, whereas the latter is due to differing vdW interaction between ground-state and excited molecules. The long-range Coulombic interaction can be further extended to 1D arrays of unit cells containing two bases, where the number of unit cells represents the degree of exciton delocalization. For example, the inter-dipole interaction will split LDS and UDS into two states for double cells and N states for N cells (Fig. 4a). Because only the uppermost (lowermost) state is optically allowed for the LDS (UDS) band,[11] the optical transition energy increases (decreases) for LDS (UDS) with increasing N. As excitons are more delocalized at a lower temperature because of reduced phonon-induced dephasing,[8] the prediction in Fig. 4a is qualitatively consistent with the differing shift rates of the DS peak energies (Fig. 3c). For a more realistic description, one may need to consider the influence of structural disorder that can be present in samples. Disorder limits the degree of delocalization and induces a redshift in excitonic states under short-range potential fluctuations.[44]

To account for the long-range Coulombic interaction among transition dipole moments of excitons delocalized in 2D crystals, we performed a simple electrostatic calculation under the point-dipole approximation.[11] Using the structural information revealed by the diffraction and polarized spectroscopy, 2D arrays of transition dipoles for $1L_P$ were generated as schematically given in Fig. 4b. The reference dipole (black arrow; basis I) at the center interacts attractively and repulsively with four type-II bases (first-shell neighbors) under LDS (red arrow) and UDS (blue arrow) coupling, respectively. Figure 4c shows that the energy for UDS (LDS) coupling decreases (increases) with increasing the number of interacting dipoles represented by the coherence length,[45] which is defined as the number of type-II shells interacting with the reference dipole in Fig. 4b. As $L_c$ is equivalent to length, it is proportional to $N_c^{1/2}$. We note that UDS is stabilized by ~10 meV when $L_c$ increases sufficiently and this value is one-third of the change in UDS energy as temperature decreased from 298 K to 77 K (Fig. 3c). The opposite $L_c$-dependences of LDS and UDS captured in Fig. 3c provides a qualitative justification for the observed energy inversion despite the limitations of the simple model. Quantitative description can be obtained by considering exciton-phonon interactions[46] and non-dipolar short-range interactions.[15]

We note that the significant redshift found for 2D PTCDA differs from the behavior of bulk polyacene crystals.[47] Negligibly small temperature dependence of the excitonic energy of the bulk system is due to the almost complete cancellation of two competing factors: thermal expansion and exciton-phonon coupling.[47] The sole contribution of the former is also insignificant for the current system assuming compliant thermal expansion: despite its negative thermal expansion coefficient (TEC),[48] few-nm-thick hBN shrinks with decreasing temperature when supported on amorphous quartz substrates with positive TEC because hBN complies with the thermal expansion of underlying substrates because of the vdW bonding between the two.[49] Similarly, 2D PTCDA is likely to comply with hBN because of their epitaxial registry. Then, the TEC of amorphous quartz predicts that the substrate-induced thermal expansion of 2D PTCDA by cooling from 298 to 79 K is less than -0.02%,[50] for which many-body calculations[47] predicted sub-meV redshift for polyacene molecular crystals. In case 2D PTCDA does not comply with hBN, however, its contraction induced by the cooling may reach up to 1% based on its bulk TEC.[51] Then, the contraction-induced redshift can be nonnegligible and may contribute to the observed redshift.

The superradiant behavior of 2D PTCDA shown in the spectral narrowing and redshift indicates that both DS excitons reside in multiple molecules, the number of which can be equated to $N_c$. Because $N_c$ is inversely proportional to $\sigma^2$,[52] it can be determined by comparing $1L_P$ with monomeric PTCDA dissolved in solvents as shown in Fig. 3f. For a similar dielectric environment as PTCDA crystals with a dielectric constant ($\varepsilon$) of 3.6,[53] chloroform ($\varepsilon = 4.81$) and toluene ($\varepsilon = 2.38$) were used as solvents. As shown in Supplementary Figure 8, monomeric PL spectra exhibited prominent vibrational progressions[32] like $1L_P$ but with twice larger linewidth. Figure 3f shows that $N_c$ increases from ~3 for 298 K to ~7 for 77 K. Viewing the simplicity of the model in Fig. 4, the estimated change in $N_c$ needs to be considered semi-quantitatively: at 298 K, a given excited molecule (basis I) interacts mainly with neighboring molecules in the first shell (basis II), which leads to DS. At a lower temperature, the excitonic wavefunction becomes more delocalized beyond the first shell. Whereas $N_c$ of ~4 and ~14 have recently been obtained at 77 K for single-basis brick-wall-type PTCDA films,[43,54] the in-plane excitonic coupling within bulk (102) plane had been considered negligible.[9]

**Out-of-plane coupling via Frenkel-CT mixing in few-layered PTCDA crystals.** In this section, we show how CT excitons are modified by mixing with FE states using multilayer PTCDA crystals. CT excitons often dominate or affect the excited state of molecular aggregates with significant π-π interactions.[19] When CT states are energetically close to Frenkel states, their coupling can be significant enough to affect their transition energies and transition dipoles. Using momentum-resolved electronic energy loss spectroscopy, for example, Knupfer et al. observed an energy-dispersive state from PTCDA crystals and derived the presence of Frenkel-CT coupling.[55] Whereas mixing-induced energetic variation is known for some other representative systems,[21, 56] the effects on transition dipoles have not been directly observed.[9] In this regard, thickness-controlled 2D PTCDA is a good model system to unravel the progression of Frenkel-CT mixing. Previous studies[9, 20, 36] established that π-π interactions are so dominant that bulk PTCDA crystals can be viewed as a non-interacting assembly of 1D π-stacked PTCDA columns (Fig. 1d). Figure 5a shows the polarized PL spectra obtained for various thickness at 77 K. Unlike 1$L_P$, thicker crystals exhibited a few broad bands including $CT_{high}$ and $CT_{low}$, which are located at slightly higher energies than 298 K (see Supplementary Figure 9 for complete angle-resolved PL spectra for 1$L_P$ ~ 6$L_P$). We note that the 0-0 emission, mostly originating from LDS at 77 K (Supplementary Figure 9), is still preserved in multilayers, whereas the other vibronic bands are not discernible because of the intense CT emissions. As shown in Supplementary Figure 9, the LDS emission of multilayers is also plane-polarized. This fact suggests that the 0-0 emission of multilayers originates from FEs localized at the surface layers[57] and subsequently not coupled to CT states. In contrast, the CT emissions exhibited a non-Malus behavior with distinct dependence on thickness. As shown in Fig. 5b~c and Supplementary Figure 9c, the polar graphs of the CT peaks for 2~3$L_P$ are elliptical unlike that of LDS, which suggests that each CT peak consists of transitions from multiple states. A similar observation could be made for 2$L_P$ ~ 6$L_P$ samples (Supplementary Figure 9). Remarkably, the transition dipole moments ($\mu_{CT}$) of CT states approach $\mu_{LDS}$ in orientation with increasing thickness (Supplementary Figure 9). Assuming that $\mu_{CT}$ is aligned along the major axis of ellipses, the angle ($\theta_{CT-LDS}$) between $\mu_{CT}$ and $\mu_{LDS}$ was determined (Fig. 5d): $\theta_{CT-LDS}$ decreased from 67° (2L) to 19° (6L).

As depicted in Fig. 5e, the thickness-dependent reorientation of $\mu_{CT}$ is attributed to the varying degree in Frenkel-CT mixing. For qualitative understanding, we adapted a perturbative mixing model[12, 58] to multilayer PTCDA crystals as given in Supplementary Note 1. In essence, observed CT states are viewed as a mixed state of unperturbed CT ($CT^o$) and FE states (Fig. 5e), and can be described in a perturbation approach.[12] For π-stacked dimers representing $2L_P$ shown in Fig. 5f, the mixing generates symmetric (+) and antisymmetric (-) wavefunctions (Equation S3 of Supplementary Note 1). Whereas the transition dipole moment ($\mu_{CT}^+$) of the former vanishes because of symmetry, that ($\mu_{CT}^-$) of the latter is modified from $\mu_{CT}^o$ by $\sqrt{2}(\kappa - \lambda) \mu_{FE}^o$, where κ and λ can be related to hole and electron transfer integrals, respectively.[58] When expanded to n-mers for $nL_P$, the degree of FE-CT mixing may decrease because of the increasing energy difference between two constituent unperturbed states (Supplementary Note 1). This suggests that the contribution of $FE^o$ to CT state decreases with increasing thickness, which is schematically depicted in Supplementary Figure 10a: For $1L_P$, $\mu_{LDS}$ and $\mu_{UDS}$ are formed orthogonal (left). With an additional layer, $\mu_{CT}^o$ aligned along ***n*** axis (3D description is given in Supplementary Figure 11) is modified by $\mu_{LDS}$ and mixed into $\mu_{CT}$ (magenta arrow; middle). For thicker layers (right), $\mu_{CT}$ is reoriented further towards ***n*** axis because of the decreased Frenkel contribution, which is consistent with the observation in Fig. 5d. One might consider the opposite case, where the contribution of $FE^o$ rises with increasing thickness (Supplementary Note 1). As depicted in Supplementary Figure 10b, $\mu_{CT}$ will eventually reorient and become more aligned with $\mu_{LDS}$ for thick layers.

We note that the model analysis not only supports the experimental observation in Fig. 5d but also provides valuable insight into the nature of the Frenkel-CT mixing. Specifically, the current results visualized the orientational progression of Frenkel-CT mixing by mapping transition dipoles with respect to crystallographic axes, which is unprecedented to our knowledge. It is also notable that the experimental $\theta_{CT-LDS}$ approached its theoretical limit of

~21.5º (Supplementary Figure 10a) with increasing thickness to 6L. This sets a rough thickness limit for the bulk behavior in regard to FE-CT mixing. Finally, thickness-dependent mixing can also be utilized in modulating exciton transport. As predicted for π-stacked dimer systems,[59] the mobility of exciton is governed by the sum of long-range Coulomb and short-range CT couplings, the latter of which is highly sensitive to the detailed geometry of intermolecular stacking. In addition to the geometric manipulation of the short-range coupling,[59] the control over FE-CT mixing suggested in the current work will enable novel functions and applications in various transport devices.

In this work, we described the spatial evolution of the intralayer and interlayer excitonic couplings in 2D PTCDA crystals using polarization-resolved variable-temperature spectroscopies. 2D PTCDA grown on hBN crystals by physical vapor deposition enabled fine controls over the thickness and crystallographic orientation. Electron diffractograms combined with polarized spectra led to the crystallographic details including molecular orientations of two herringbone-arranged bases. For 1L, the 0-0 emission pair Davydov-split by Kasha-type coupling exhibited a remarkable energy inversion with decreasing temperature, which favors in-plane delocalization of the Frenkel exciton pair. With increasing thickness, CT excitons developed in the out-of-plane direction along π-π stacks. Notably, the transition dipoles of CT excitons underwent in-plane reorientation with increasing thickness, which was successfully explained with a Frenkel-CT mixing model. This work will lead to a deeper understanding of molecular excitons bound in low-dimensional organic materials and thus contribute to the better design of molecular structures with novel functions.

**Methods**

**Growth of 2D PTCDA crystals.** Mono- and few-layer PTCDA crystals were grown on high-quality inorganic 2D crystals by the self-limited epitaxy.[23] Briefly, few-layer hBN and graphene were mechanically exfoliated on amorphous quartz and 285 nm $SiO_2$/Si substrates from hBN powder (Momentive Performance Materials) and natural graphite (NGS Naturgraphite GmbH). PTCDA powder (TCI, >98%) and exfoliated samples were placed at the center and downstream of a tube furnace (ThermoFisher, Lindberg/Blue M), respectively. After the quartz tube was purged with a high-purity Ar gas for 30 min at 500 mL/min, the flow rate was reduced to 100 mL/min. Subsequently, the tube was heated to 330 °C at a rate of ~22 °C/min and the temperature was maintained for 15 ~ 30 min. Sublimated PTCDA molecules were found to deposit as 2D crystals on hBN and graphene. The average thickness was controlled by varying the deposition time and location within the tube.

**PL spectroscopy.** The employed micro-spectroscopy setup has been described elsewhere.[60] Briefly, solid-state laser beams of 457.8 and 514.3 nm (Cobolt, Twist and Fandango) were focused onto samples within a spot size of ~1 μm using a microscope objective (40X, NA = 0.60). The signals were collected using a CCD camera (Princeton Instruments, PyLon) with a spectrometer (Princeton Instruments, SP2300). All measurements were performed in the ambient conditions unless noted otherwise. The average power was maintained below 2 μW to avoid unwanted photoinduced degradation.

**Temperature-dependent PL spectroscopy.** A tunable pulsed source (Coherent, Chameleon Ultra II and Compact OPO) was used for variable-temperature PL measurements. Excitation beams of 515 nm were focused with an objective (50X, NA = 0.55) onto a 1-μm focal spot on samples. The light source was operated at a repetition rate of 80 MHz generating pulses with 140-fs duration. Samples placed in a liquid-$N_2$ cryostat (Linkam, HFS600E-PB4) could be cooled down to 77 K. PL signals were collected by a Czerny-Turner spectrometer (Andor, SR303i) equipped with an EMCCD camera (Andor, DU971P). For polarization-resolved experiments, an analyzing polarizer was placed in front of the spectrometer to select the polarization component parallel to that of the plane-polarized excitation beam. The polarization vector of the excitation beam could be rotated within the sample plane by rotating an achromatic half-wave plate, which was mounted on a rotational stage (Thorlabs, KPRM1E/M) and placed between the objective and tube lens. The average power was maintained below 0.2 μW to avoid unwanted photoinduced degradation.

**Optical contrast and differential reflectance.** Optical absorption of 2D PTCDA on transparent substrates could be measured using its differential reflectance (DR),[35] defined as $(R - R_o)/R_o$, where $R$ and $R_o$ are the spectral reflectance of sample and bare substrate, respectively. Optical contrast, essentially DR, was similarly determined using the green-channel signals of optical micrographs.

**AFM measurements.** The topographic details of samples were studied with an atomic force microscope (Park Systems, XE-70). Height and phase images were obtained in a noncontact mode using Si tips with a nominal radius of 8 nm (MicroMasch, NSC-15).

**TEM sample preparation and measurements.** PTCDA on 2D crystals (hBN or graphene) was covered with few-layer graphene flakes before transferring to TEM grids. Then, the resulting samples of graphene/PTCDA/hBN (or graphene)/SiO$_2$/Si were overlaid with polydimethylsiloxane (PDMS) film. When soaked in ethanol/DI water mixtures (ethanol 50%, DI water 50%) for 2 h, the 2D hetero layers of the samples were released from SiO$_2$/Si substrates and transferred to the PDMS film. The heterostructures on PDMS were dried under ambient conditions and transferred again to holey SiN TEM grids (Norcada Inc.) by stamping via a micromanipulator under optical microscope observation. Selected area electron diffraction (SAED) acquisition was performed with a JEOL ARM 200F equipped with image and probe aberration correctors operated at 80 kV. SAED simulations were performed by CrystalDiffract (CrystalMaker Software Ltd.).

**Data availability**

The authors declare that all data supporting the findings of this study are available within the paper and its supplementary information file or available from the corresponding author upon request.

**ACKNOWLEDGMENT**

S.R. acknowledges the financial support from Samsung Research Funding Center of Samsung Electronics (Project Number SSTF-BA1702-08), National Research Foundation of Korea (NRF-2020R1A2C2004865 and NRF-2022R1A4A1033247) and Samsung Electronics Co., Ltd (IO201215-08191-01).

**Author Contributions**

S.R. conceived the project. D.K. and S.R. designed the experiments. D.K. performed the spectroscopy experiments and analyzed the data. S.L. and K.K. conducted the SAED measurement and analysis. D.K. and S.R. wrote the manuscript with contributions from all authors.

**Notes**

The authors declare no conflict of interest.

**Figures and Captions**

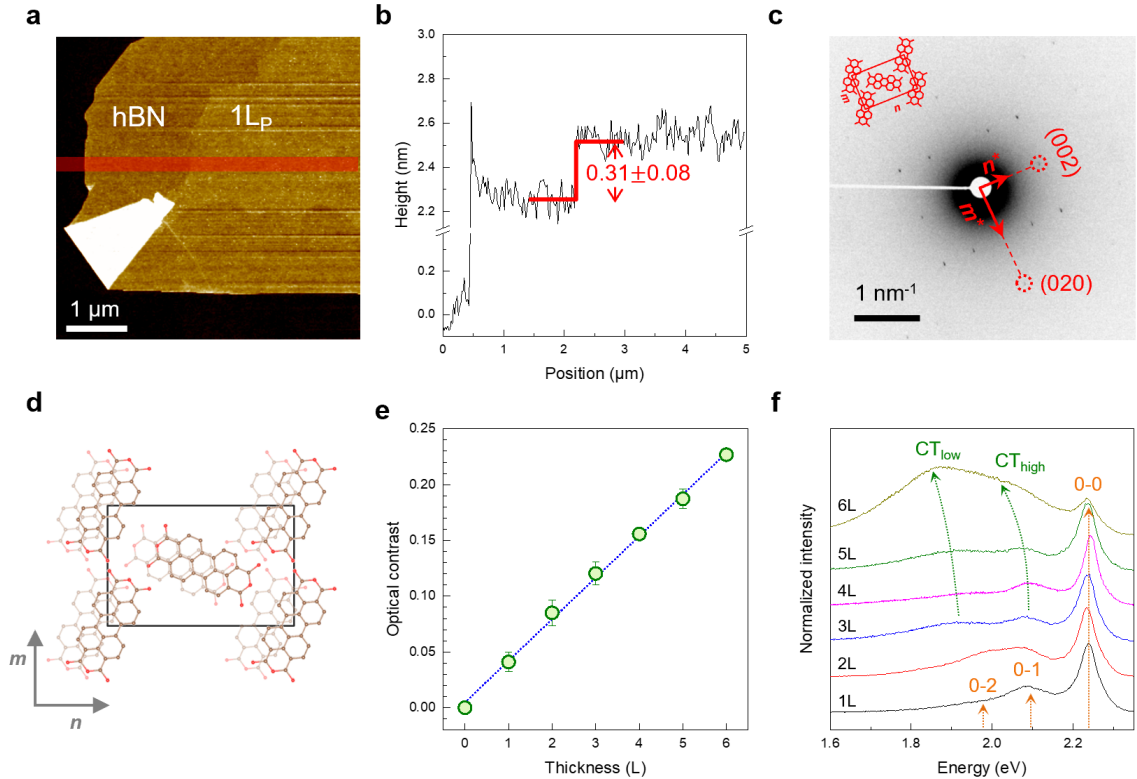

**Figure 1. Structural and optical characterization of 2D PTCDA crystals.** (a) AFM height image of $1L_P$ on few-layer hBN/SiO$_2$/Si (see Supplementary Figure 1 for its optical micrographs before and after PVD). (b) Averaged height profile obtained in the red area in (a). (c) SAED image from $2L_G/1L_P/nL_{BN}$, where $2L_G$ served as a protection layer (see Methods and Supplementary Figure 12 for details). The optical micrograph of the assembled heterostructure is given in Supplementary Figure 2. The inset presents a schematic two-basis unit cell of $1L_P$ crystals determined by electron diffraction and polarized PL spectroscopy, where unit vectors ***m*** and ***n*** correspond to ***b*** and $[\bar{2}01]$ of bulk crystals. The real-space orientation of the unit cell, rectangle in (d), was determined by the reciprocal vectors (***m**** and ***n****) of diffraction images. (d) Proposed structure of $2L_P$ crystals resembling α-type polymorph with two basis molecules in a herringbone configuration. The molecules in pale colors belong to the bottom layer. (e) Optical contrast of $1L_P \sim 6L_P$ on hBN/quartz determined using the green-channel signals of optical micrographs. The dashed line is a linear fit to the data. Error

bars indicate standard deviations obtained from multiple samples. (f) Unpolarized PL spectra of $1L_P \sim 6L_P$ on hBN. See the text for the assignment of various PL peaks.

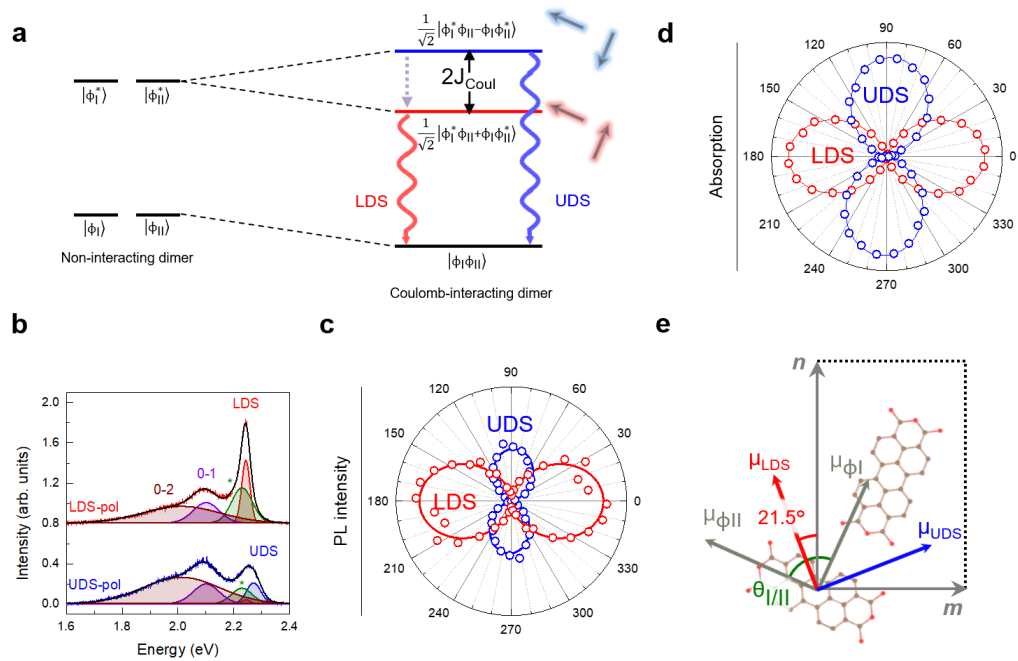

**Figure 2. Optical anisotropy and epitaxial growth of 1L PTCDA.** (a) Scheme of Davydov splitting in an oblique PTCDA dimer described by ground ($\varphi_i$) and excited-state wavefunctions ($\varphi_i^*$). LDS and UDS are the superpositions of $\varphi_i$ and $\varphi_i^*$ resulting from the Coulombic interactions ($J_{Coul}$) between molecular transition dipoles (oblique arrows). (b) LDS and UDS-polarized PL spectra of $1L_P$ on hBN obtained in a parallel configuration, where the polarizations of the excitation beam and PL signals were aligned with $\mu_{LDS}$ (upper) and $\mu_{UDS}$ (lower), respectively. Complete sets of spectra (Supplementary Figure 5) were globally fitted with multiple Gaussian functions for 0-v vibronic components, where v refers to the vibrational quantum number of the electronic ground state and the pair of LDS and UDS corresponds to the 0-0 transition. The origin of the peak at 2.21 eV (marked with asterisks) is unclear. (c) Polar graph of PL intensity for LDS and UDS peaks, where the data were fitted with the square of cosine. (d) Polar graph of differential reflectance (DR) for LDS and UDS peaks, for the latter of which an angle-independent constant was subtracted (see Supplementary Figure 2 for complete DR spectra). (e) Orientations of molecular and DS transition dipole moments determined with polarized spectroscopy and electron diffraction.

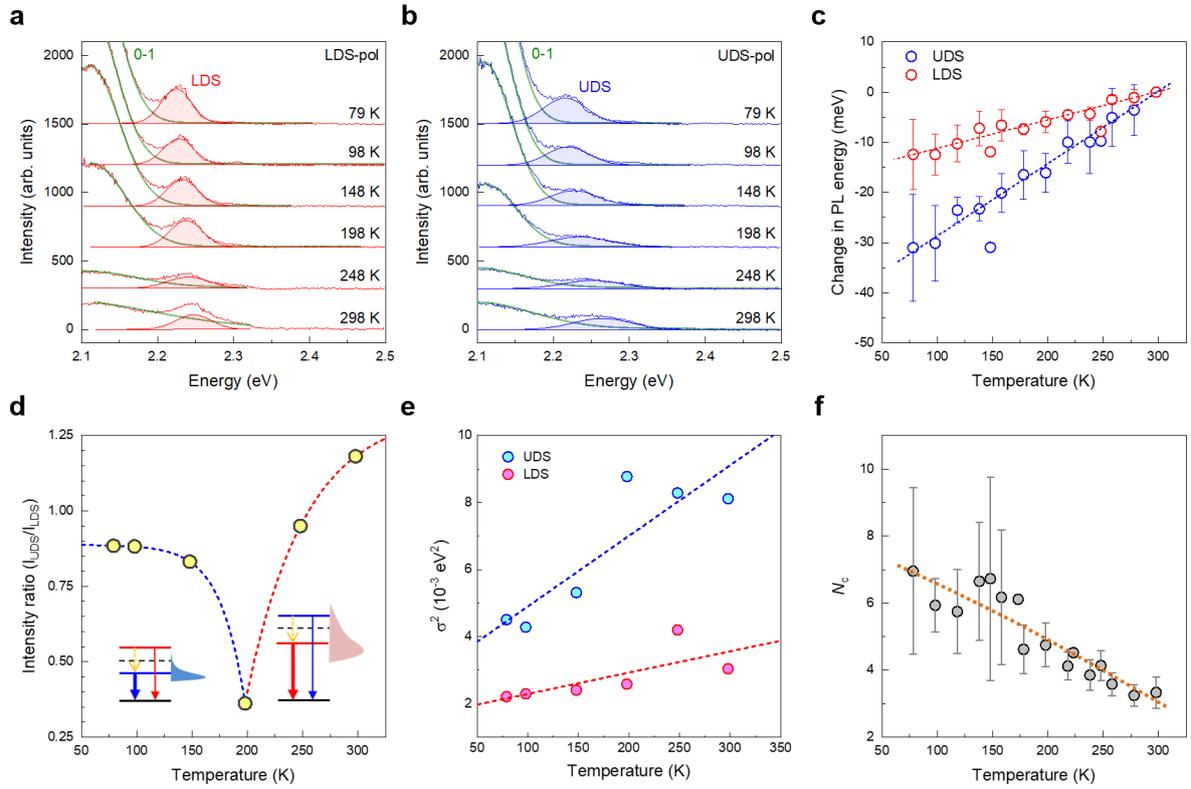

**Figure 3. Temperature-dependent Davydov splitting.** (a & b) Polarized PL spectra of 1L$_P$/hBN: LDS- (a) and UDS-polarized (b) spectra obtained for 79 ~ 298 K. Each spectrum in (a ~ b) was fitted with two Gaussian functions for 0-0 and 0-1 peaks. Note that the unknown peak at 2.2 eV identified in Fig. 2b was not included in the fitting because of the lack of sufficient spectra required for a global analysis. (c) Change in PL energy when referenced at 298 K. (d & e) Intensity ratio (d) and FWHM squared, $\sigma^2$, (e) of LDS and UDS peaks. The scheme in (d) depicts an energy inversion in DS and the associated change in Boltzmann-type populations with decreasing temperature. (f) Coherence number ($N_c$) of 1L$_P$ determined by comparing linewidths following: $N_c = \sigma_{mono}^2/\sigma_{2D}^2$ and $N_c \propto 1/T$,[52] where $\sigma_{mono}$ ($\sigma_{2D}$) is FWHM of 0-0 peaks for monomeric (2D) PTCDA and T is temperature, respectively. PL spectra of monomeric solutions are given in Supplementary Figure 8. Dotted lines in (c) and (e) are linear fits, whereas those in (d) and (f) are guides to the eye. Error bars in (c) and (f) indicate standard deviations obtained from multiple samples.

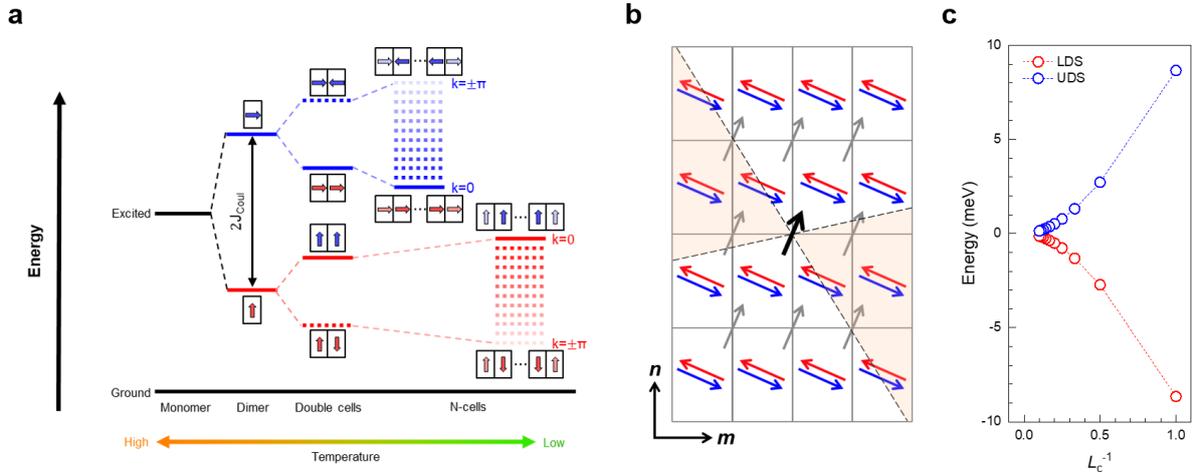

**Figure 4. Extended 2D excitonic coherence at low temperature.** (a) Excitonic progression in 1D molecular crystals with increasing coherence. For a dimer represented by a unit cell (black rectangle), the Coulomb interaction ($J_{Coul}$) between two molecular bases (I and II) leads to the LDS (red arrows) and UDS (blue arrows) dipoles. For simplicity, the LDS and UDS dipoles are set parallel to one of the unit cell edges. With decreasing temperature, more DS dipoles undergo coherent coupling, which can be approximated as the double-cell and N-cell cases in (a). The delocalized coupling leads to excitonic bands for LDS (red dotted band) and UDS (blue dotted band). Only the uppermost (lowermost) state is optically accessible for the LDS (UDS) band, where $k$ represents the crystal momentum. (b) 2D arrangement of molecular transition dipoles determined for $1L_P$ and used for the electrostatic calculations shown in (c). Four type-II dipoles (red and blue arrows) in the first shell of the reference dipole at the center (black arrow; type-I) orient in opposite directions for LDS (red arrows) and UDS (blue arrows) coupling, whereas all type-I dipoles including eight in the second shell (gray arrows) remain unchanged. Type-I dipoles in the two orange quadrants interact repulsively with the reference dipole. (c) Energy of the reference transition dipole moment interacting with another delocalized over neighboring type-II dipoles. Coherence length[45] is defined as the number of type-II shells surrounding the reference dipole: the first and second shells contain 4 and 12 type-2 dipoles, respectively. We used 6.93 Debye for the monomeric transition dipole moment[61] and $\varepsilon = 1$ for simplicity.

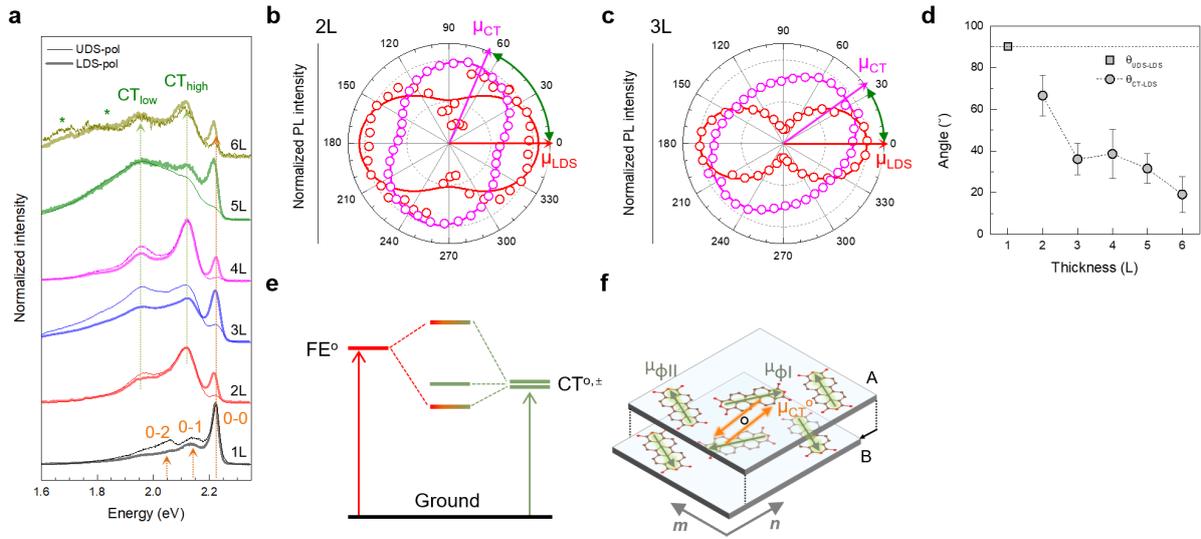

**Figure 5. Spatial reorientation of CT excitons with varying thickness.** (a) LDS- and UDS-polarized PL spectra of nL$_P$ on hBN obtained at 77 K. For the asterisk-marked peaks below 1.9 eV, see Supplementary Figure 9. (b & c) Angle-resolved PL intensity of LDS and CT peaks from 2L$_P$ (b) and 3L$_P$ (c), where the data were fitted with the square of cosine and a constant. The angle marked by a green arrow corresponds to θ$_{CT-LDS}$ in (d). (d) Orientation of μ$_{CT}$ and μ$_{UDS}$ with respect to μ$_{LDS}$. (e) Energy level diagram for FE-CT mixing in 2D PTCDA crystals. Whereas LDS (2.23 eV) for 1L$_P$ can be considered as an unperturbed FE state (FE$^o$), the CT state at 2.12 eV from multilayer samples is the result of FE-CT mixing. For the symmetry convention of unperturbed CT states (CT$^{o,\pm}$), see the main text. (f) Scheme presenting the transition dipole moments of FE$^o$ (green arrows) and CT$^o$ (orange arrows) states for 2L$_P$. Because of its crystallographic centrosymmetry about the black circle, the two neighboring μ$_{φI}$ dipoles in the upper (A) and lower (B) (102)-planes are antiparallel to each other. They need to be combined in parallel for a dipole-allowed transition. Similarly, the two CT$^o$ dipoles for electron and hole transfer ($\vec{\mu}_{A^+B^-}{}^o$ and $\vec{\mu}_{A^-B^+}{}^o$) require a parallel combination for a bright state. See the main text and Supplementary Note 1 for details.